\documentclass[10pt, aps, prb, twocolumn,amsmath,amssymb,showpacs,groupedaddress]{revtex4-1}
\usepackage{graphicx}
\usepackage{comment}
\begin{document}

\preprint{}


\renewcommand{\vec}[1]{\mathbf{#1}}

\title{Lattice dynamics and antiferroelectricity in PbZrO$_3$ tested by X-ray and Brillouin light scattering}
\author{R. G. Burkovsky}
\email{burkovsk@esrf.fr}
\affiliation{European Synchrotron Radiation Facility, F-38043 Grenoble Cedex, France}
\affiliation{St.Petersburg State Polytechnical University, 29 Politekhnicheskaya, 195251, St.-Petersburg, Russia}

\author{A. K. Tagantsev}
\affiliation{Ceramics Laboratory, Swiss Federal Institute of Technology (EPFL), CH-1015 Lausanne, Switzerland}
\affiliation{Ioffe Institute, 26 Politekhnicheskaya, 194021, St.-Petersburg, Russia}

\author{K. Vaideeswaran}
\affiliation{Ceramics Laboratory, Swiss Federal Institute of Technology (EPFL), CH-1015 Lausanne, Switzerland}

\author{S. B. Vakhrushev}
\affiliation{Ioffe Institute, 26 Politekhnicheskaya, 194021, St.-Petersburg, Russia}
\affiliation{St.Petersburg State Polytechnical University, 29 Politekhnicheskaya, 195251, St.-Petersburg, Russia}

\author{A. V. Filimonov}
\affiliation{St.Petersburg State Polytechnical University, 29 Politekhnicheskaya, 195251, St.-Petersburg, Russia}

\author{A. Shaganov}
\affiliation{St.Petersburg State Polytechnical University, 29 Politekhnicheskaya, 195251, St.-Petersburg, Russia}

\author{D. Andronikova}
\affiliation{St.Petersburg State Polytechnical University, 29 Politekhnicheskaya, 195251, St.-Petersburg, Russia}

\author{A. I. Rudskoy}
\affiliation{St.Petersburg State Polytechnical University, 29 Politekhnicheskaya, 195251, St.-Petersburg, Russia}

\author{A. Q. R. Baron}
\affiliation{Materials Dynamics Laboratory, RIKEN SPring-8 Center, 1-1-1 Kouto, Sayo 679-5148 JAPAN}

\author{H. Uchiyama}
\affiliation{Research and Utilization Division, SPring-8/JASRI, 1-1-1 Kouto, Sayo 679-5198 JAPAN}

\author{D. Chernyshov}
\affiliation{Swiss-Norwegian Beamlines at ESRF, F-38043 Grenoble Cedex, France}


\author{Z. Ujma}
\affiliation{Institute of Physics, University of Silesia, ul. Uniwersytecka 4, 40-007 Katowice, Poland}

\author{K. Roleder}
\affiliation{Institute of Physics, University of Silesia, ul. Uniwersytecka 4, 40-007 Katowice, Poland}

\author{A. Majchrowski}
\affiliation{Institute of Applied Physics, Military University of Technology, ul. Kaliskiego 2, 00-908 Warsaw, Poland}

\author{Jae-Hyeon Ko}
\affiliation{Department of Physics, Hallym University, 39 Hallymdaehakgil, Chuncheon, Gangwondo 200-702, Korea}

\author{N. Setter}
\affiliation{Ceramics Laboratory, Swiss Federal Institute of Technology (EPFL), CH-1015 Lausanne, Switzerland}

\date{\today}

\begin{abstract}
We report the results of comprehensive study of the critical dynamics of the prototype perovskite antiferroelectric PbZrO$_3$. The combination of   inelastic X-ray and diffuse X-ray scattering techniques and Brillouin light scattering was used. It is found that the dispersion of the TA phonons is strongly anisotropic. The dispersion curve of the in-plane polarized TA phonons propagating in [1~1~0] direction demonstrates pronounced softening. Slowing down of the excitations at R-point is found, it is manifested in growing of the central peak. This slowing down is too weak to be considered as a primary origin of the corresponding order parameter.  Obtained results are treated in terms of TA--TO flexoelectric mode coupling. It is demonstrated that the structural phase transformation in PbZrO$_3$ can be considered as the result of the only intrinsic instability associated with the ferroelectric soft mode.

\end{abstract}


\maketitle


\section{Introduction}
Antiferroelectricity lies at the roots of already widely used high-performance functional materials like PZT and may become even more important in context of the new ideas about its practical applications \cite{Wei2014}. Both the phenomenon itself and its underlying mechanisms are currently under intense investigation and review \cite{Bussmann2013,Ko2013,Tagantsev2013,Rabe2013,Hlinka2014,Hao2014}. 
Kittel \cite{kittel1951} initially assigned the term ``antiferroelectric'' to the crystal that has a structural phase transition from a higher-symmetry nonpolar parent phase to a lower-symmetry nonpolar phase, distinguished from the parent phase by the anti-parallel ionic shifts. On approaching the transition temperature experimentally one typically sees an increase of the dielectric constant, $\varepsilon(T)$. In prototypical antiferroelectric PbZrO$_3$ the growth of $\varepsilon(T)$ is remarkable and resembles the critical growth of $\varepsilon(T)$ in ferroelectrics. This suggests that ferroelectrics and antiferroelectrics share the same type of ferroelectric instability associated with softening of the transverse optic mode in the high-symmetry phase. 
Kittel's model does not require critical growth of $\varepsilon(T)$ and instead admits a finite, not necessarily large, value of $\varepsilon(T)$ close to the transition temperature. 
However important insights into the physical picture behind antiferroelectricity were obtained with models having explicitly two instabilities: one ferroelectric-like instability associated with the soft zone-center TO mode producing the dielectric anomaly and a second one driving structural changes \cite{Cross1956, Okada1969, Balashova1993}. The two-instability model allows one to describe the key features of an antiferroelectric: the specific temperature dependence of the dielectric permittivity accompanied by a phase transition between the two non-polar phases and the double-hysteresis loops.
The assumption of the anti-polar character of displacements is not strictly necessary in this type of models. In principle the structural instability may be associated with any type of lattice distortions provided that the corresponding order parameter is linked to the ferroelectric order parameter by the proper type of biquadratic coupling \cite{Balashova1993}. 

Two instabilities are not enough for explaining the structural changes in PbZrO$_3$\ (PZO). 
In this crystal the low-temperature structural distortion with respect to the parent cubic phase 
is determined by two order parameters with different symmetries.
One of these is desribed by the wavevector $\vec{q_{\Sigma}}$ =(1/4 1/4 0) and corresponds to the antiparallel shifts of the lead ions while the other is described by the wavevector $\vec{q_R}$ =(1/2 1/2 1/2) and corresponds to the anti-phase tilts of oxygen octahedra \cite{Whatmore1979} where the components of $\vec{q}$ are measured in reciprocal lattice units, $2\pi/a$, where $a$ is a cubic lattice constant. 
Consequently, simultaneous critical increase of 3 generalized susceptibilities might be expected in the high-symmetry phase on approaching the transition temperature. 
Until recently \cite{Tagantsev2013} no observations in favour of or against such critical behavior  were reported, except for the dielectric susceptibility\cite{Shirane1951} that demonstrates the critical divergence with Curie temperature $T_c=$485~K.
 
Direct ways to measure the generalized susceptibility away from the Brillouin zone center include neutron and X-ray inelastic scattering. Some qualitative information can be also extracted from energy-integrated diffuse scattering experiments.  
Small size of presently available PbZrO$_3$\ (PZO) single crystal prevents using the neutron scattering technique for studying non-Bragg scattering, leaving this task to the X-ray techniques. Recently we used the latter to show that the AFE transition in PbZrO$_3$ may be interpreted as an incommensurate phase transition going directly to a lock-in phase \cite{Tagantsev2013}. In this paper we give a comprehensive report on the critical dynamics in PZO as evidenced by complementary use of inelastic X-ray scattering, diffuse X-ray scattering and Brillouin light scattering techniques. We show that the only generalized susceptibility that tends to diverge near the transition temperature is the one associated with the softening transverse optic (TO) mode, but the primary order parameter of the transition is linked to the markedly temperature dependent transverse acoustic (TA) branch, whose temperature dependence is conditioned upon the flexoelectric interaction involving the TO mode with critical temperature dependence. The $R$-point mode shows a moderate, non-pronounced temperature dependence and the establishment of corresponding structural distortion below the transition temperature is created due to biquadratic coupling to the primary order parameter.

\section{Experimental}
Lead zirconate single crystals were grown from high temperature solutions (flux growth method) by means of spontaneous crystallization.
The Pb$_3$O$_4$-B$_2$O$_3$ mixture (soaking at 1350K) was used as a solvent. The temperature of the melt was reduced at a rate 3.5 K/h down to 1120 K.
The remaining melt was decanted and as-grown crystals attached to the crucible walls were cooled to room temperature at a rate of 10 K/h.
In the final step the as-grown crystals were etched in dilute acetic acid to remove residues of the solidified flux. The samples for the X-ray scattering experiments were taken from the same batch while the Brillouin measurements were carried out with separately grown crystals.

Studied crystals  exhibited a narrow temperature range of existence of the intermediate phase on heating with a lattice distortion corresponding to the M point ($\vec{q}_M$=(1/2 1/2 0)) of the Brillouin zone, which had been earlier reported by Fujishita and Hoshino \cite{Fujishita1984}. Tentatively it may correspond to the intermediate phase reported by Roleder et al. \cite{Roleder1988}. 

The IXS experiments were carried out using the spectrometer installed at BL35XU  \cite{Baron2000} of the SPring-8 synchrotron source in Japan. 
It was operated at X-ray energy E=21.75~keV (Si (11~11~11) monochromator) providing the energy resolution of 1.5 meV FWHM. 
We used a 20 micron thick PZO platelet as a sample.
We used the multianalyzer array of BL35XU \cite{Baron2008}. The IXS spectra were recorded simultaneously by 12 independent analyzer-detector pairs, allowing to measure the phonon resonances at 12 distinct points in reciprocal space. 
This technique enabled us to determine phonon dispersion surfaces instead of phonon dispersion curves in the high symmetry directions. Close to the reciprocal lattice point $\vec{Q}=(3\ 0\ 0)$, where Q is measured in reciprocal lattice units of PZO, the Q-resolution was $\Delta \vec{Q} = (0.06\ 0.01\ 0.06)$.

Measurements of diffuse scattering were performed at the  general purpose KUMA6 diffractometer at BM01A Swiss Norwegian Beam Line of ESRF.
A sagittally focusing Si(111) monochromator was used and wavelength $\lambda$=0.99\AA\ was selected and calibrated with the NIST LaB$_6$\ standard.
Diffraction patterns were recorded using the MAR345 detector.
All the measurements were performed with a small single crystal having the shape of a rectangular parallelepipedon of about 20x20x500~$\mu$m$^3$.
The sample was mounted at the goniometer and heated by a flow of hot nitrogen.
The temperature was regulated with an ESRF heat blower with a stability of 0.5~K. Three-dimensional reconstructions of the scattering intensity in reciprocal space and 2D cross-sections of these reconstructions were performed using Volvox program.

A conventional tandem multi-pass Fabry-Perot interferometer was used to measure the Brillouin spectrum in a narrow (±13 GHz) frequency range by using a free spectra range of 15 GHz.
The PZO sample was placed in a cryostat (FTIR 600, Linkam) which was set up vertically for forward, symmetric scattering experiment.
A solid state laser (Excelsior 532-300, SpectraPhysics) at a wavelength of 532 nm was used as an excitation source.
The details of the Brillouin spectrometer can be found elsewhere \cite{Kim2012}.


\section{Phonon dispersion}
\begin{figure*}
\includegraphics [width=.95\textwidth,clip=false, trim=0mm 0mm 0mm 0mm] {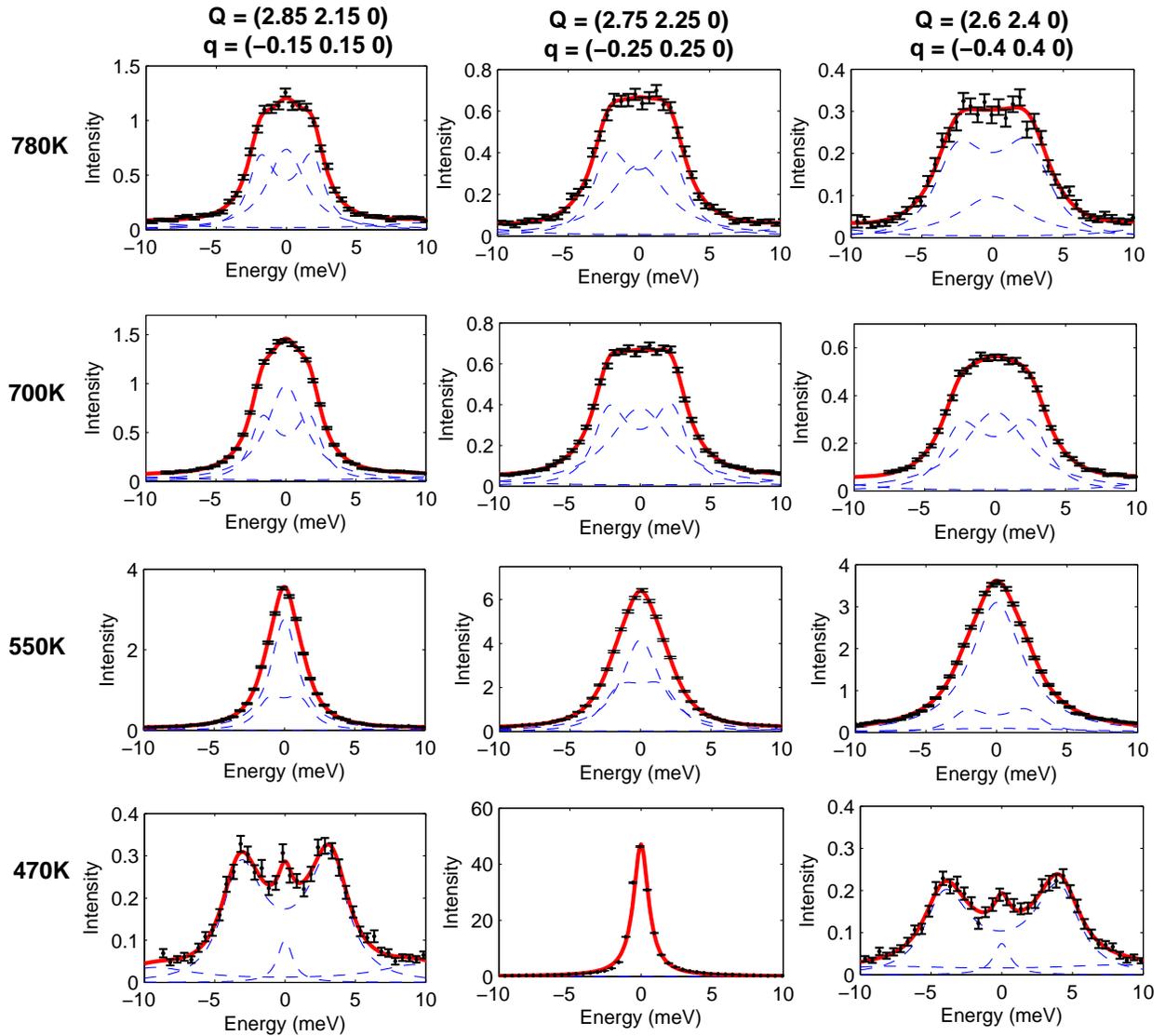}
\caption{\label{fig_gmspectra}
Temperature evolution of the IXS spectra along $\Gamma$-M direction. 
The figures are arranged in the matrix with different columns corresponding to different $q$ values and different rows to different temperatures. Dashed lines define the fit by the sum of 2 damped harmonic oscillator (DHO) and the central peak. At $T$=470 K, $q$=(0.25 0.25 0) the huge peak is observed due to formation of the superstructure.
}
\end{figure*}

\begin{figure}
\includegraphics [width=1\columnwidth,clip=true, trim=0mm 0mm 0mm 0mm] {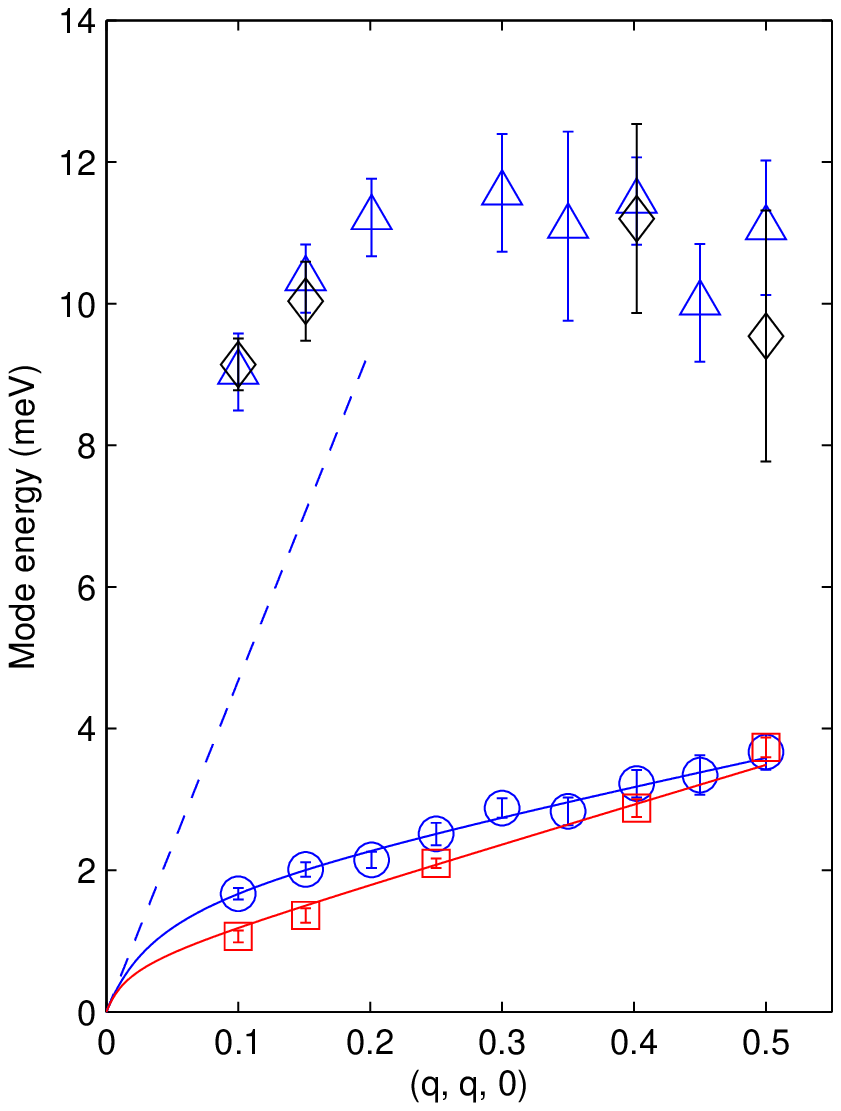}
\caption{\label{fig_dispersion} 
Dispersion of the TA and TO phonons, propagating in $\Sigma$ direction ($\vec{q} = (q,q,0)$) of the reciprocal space and polarized in (0 0 1) plane: circles -- TA at 780~K, squares -- TA at 550~K, triangles -- TO at 780~K, diamonds -- TO at 470~K (X-ray scattering data). Error bars correspond to 95\% confidence interval.
The dashed line indicates the slope of the dispersion curves in the vicinity of the $\Gamma$ point, extracted from Brillouin light scattering data. 
}
\end{figure}

In our measurements we concentrated on $q$ along [1 1 0] direction covering $\Sigma$-point and potentially interesting M-point. 
In addition we carefully studied phonons at and around R-point.
Figure \ref{fig_gmspectra} shows the spectra corresponding to the $\vec{q}$ at $\Gamma$-M line, measured in a way that mostly the transverse in-plane polarized phonons are visible. 
Hereafter by in-plane polarized phonons we refer to phonons propagating in $[c\ s\ 0]$ directions with the polarization vectors being exactly or close to $[-s\ c\ 0]$ directions where $c^2+s^2=1$. 
The spectra cover the range of points from $\vec{q}=$(0.1 0.1 0) through $\Sigma$\ (1/4 1/4 0) to $M$ (1/2 1/2 0) point.
The spectra contain the two pairs of phonon resonances and the central peak. 
Phonon resonances were fitted to damped harmonic oscillators lineshape function convoluted with the experimental resolution. 
The low energy phonon resonances can be associated with the transverse acoustic (TA) branch while the high-energy resonances should be attributed to the transverse optic (TO) branch. 
The spectra in Fig. \ref{fig_gmspectra}  show a distinct temperature dependence both in the TA phonon frequency and in the intensity of the central peak. 
At low $q$ values they become poorly distinguishable close to the transition temperautre. 
Nevertheless in a broad $T-q$ range they can be reliably analyzed separately.  Figure \ref{fig_dispersion} shows the phonon dispersion curves for the in-plane polarized TA and TO modes. 
Already at $q=$0.1 the TA phonon frequencies both at 780~K and 550~K are much lower then the value extrapolated from the sound velocity determined at 550~K (see below).  For the $q$-range from 0.1 to 0.5 the TA mode at 780~K and at 550~K has close to linear dispersion and the dispersion softens when temperature decreases. For $\vec{q}$=(0.1 0.1 0) the decrease is of about 40\% from T=780 K to T=550 K. At T=550 K the phonon energy at $\Sigma$-point in PZO is about 2 meV. 
Below the transition temperature (see T=470 K in Fig. \ref{fig_gmspectra}) the values of the in-plane TA phonon energy are recovered to about  4--5~meV typical for the lead based perovskites. This can be seen in the bottom row of Fig. \ref{fig_gmspectra} which shows the same $\Gamma$-M direction but below the transition temperature. One may note the appearance of the $\Sigma$-point superstructure in that row (note the difference in the intensity scale). The central peak, which is not related to the superstructure, is strongly suppressed below T$_c$.

\begin{figure}
\includegraphics [width=1\columnwidth,clip=true, trim=0mm 0mm 0mm 0mm] {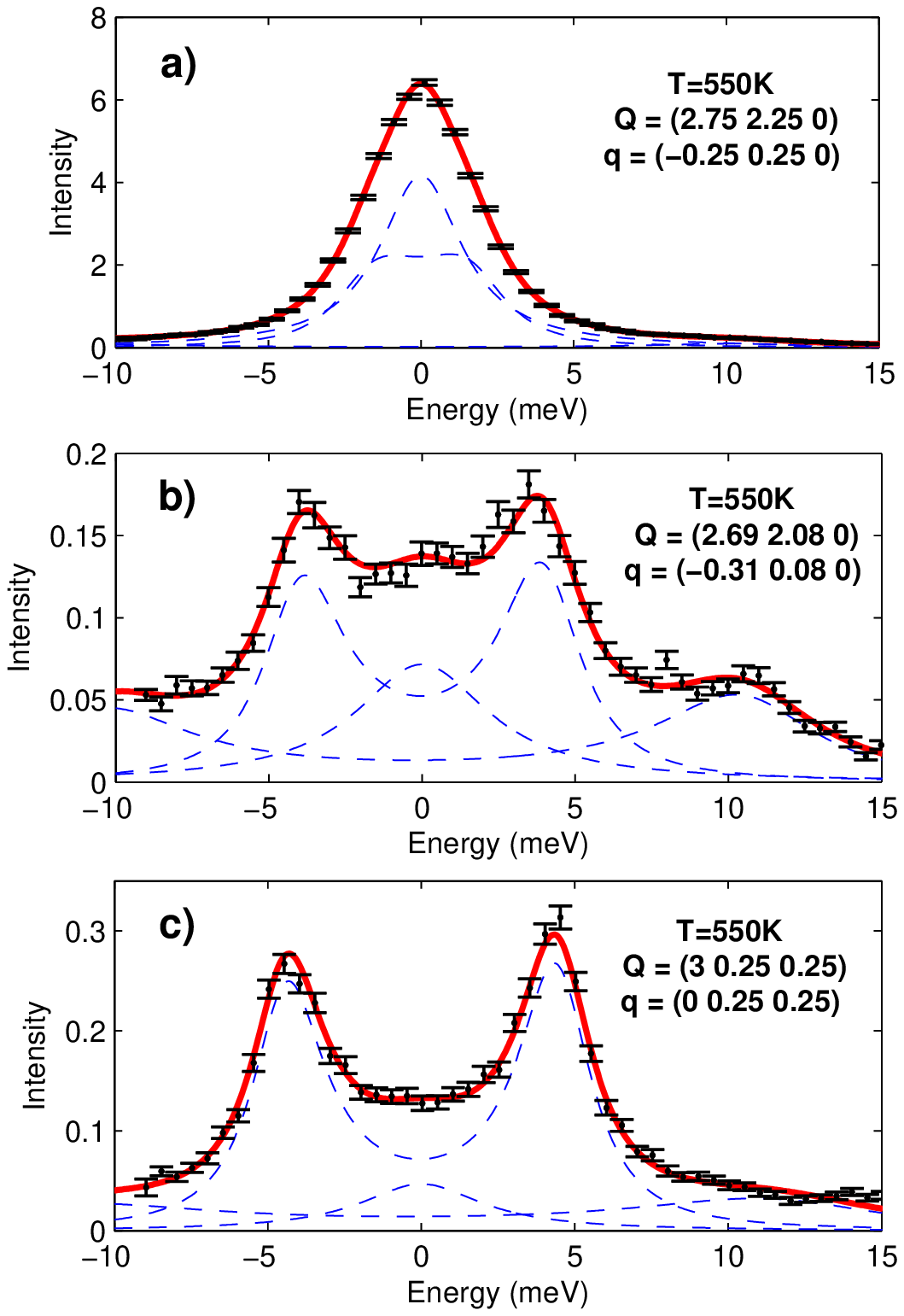}
\caption{\label{fig_dir_diff}
IXS spectra at T=550 K (E -- Energy transfer in meV) for different directions of the reduced wavevector $\vec{q}$ and different polarization. (a) $\vec{q}$ along [1 1 0] and polarization along [1 -1 0]; (b) $\vec{q}$ aside of [1 1 0] direction and polarization in HK0 plane; (c) $\vec{q}$ along [1 1 0] and polarization along [0 0 1] direction. 
}
\end{figure}

The anomalously low frequency of the TA phonon is seen only for in-plane polarized  TA phonons propagating in  $\Gamma$-M direction. Once the wavevector deviates from this direction the dispersion starts to appear more usual. The difference between different directions can be seen in Fig. \ref{fig_dir_diff} where IXS spectra for $q$\ along [1 1 0] and $q$ deviating from this direction are shown. The length of $q$ is nearly the same for different plots in the figure. It is clearly seen that moving away from the 'soft' [1 1 0] direction results both in increase of the phonon frequency and suppression of the central peak (Fig. \ref{fig_dir_diff}(b)). Also the normally-high energy is observed for the out-of-plane phonons, even for $\Gamma$-M direction (Fig. \ref{fig_dir_diff}(c)). 
The high-energy phonon resonances, observed in the spectra of Figs. \ref{fig_gmspectra} and \ref{fig_dir_diff}  and corresponding to the transverse optic (TO) mode, do not show any temperature dependence. Figure \ref{fig_dispersion} shows the dispersion of the TO phonon resonance for the temperatures $T=780$ K and $T=470$ K (below the transition). Obviously the data points above and below the transition correspond to the same curve with the precision of error bars.

\begin{figure}
\includegraphics [width=.8\columnwidth,clip=true, trim=0mm 0mm 0mm 0mm] {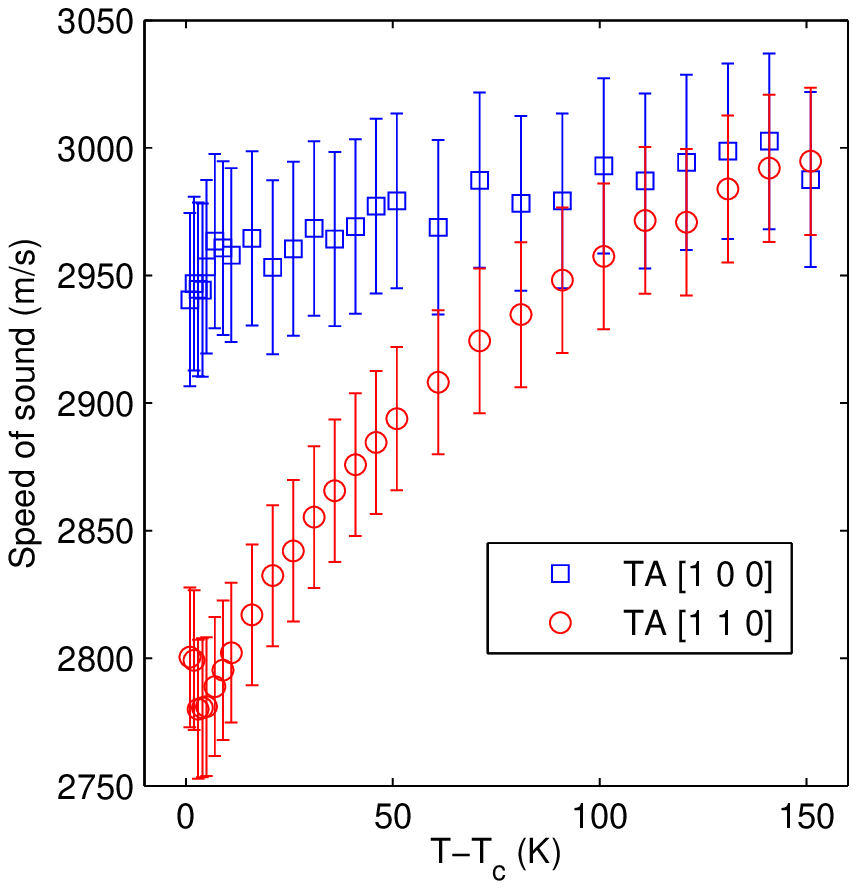}
\caption{\label{fig_brillouin}
Temperature dependence above the antiferroelectric phase transition of the speed of sound for the TA phonons in PZO determined by the Brillouin light scattering. TA [1 0 0] indicates the TA phonons propagating along the [1 0 0] direction polarized in the (1 0 0) plane, while TA[1 1 0] denotes the TA phonons propagating along the [1 1 0] direction polarized along [1 -1 0] direction. }
\end{figure}

Due to the resolution limitations we were able to observe by the IXS technique the TA resonances only with $q$-vector magnitude above or equal to 0.1 r.l.u. and these data give the dispersion relations depicted by data points in Fig. \ref{fig_dispersion}. On the other hand, at small $q$ the TA branch has a dispersion of linear form $\omega_\text{TA}=Cq$, where the constant $C$ depends on the direction and is determined by the elastic tensor of the crystal. We used Brillouin light scattering to determine the initial slopes of the dispersion curves for the acoustic branches. The temperature dependences of the speed of sound for the TA phonons propagating in [1 0 0] and [1 1 0] directions are shown in Fig. \ref{fig_brillouin}. There is about 14\% decrease in the speed of sound for the TA phonons propagating in [1 1 0] direction on cooling from approximately 660 K to the transition temperature, while for [1 0 0] direction the speed of sound does not change notably. We attribute this decrease to the effect of electrostrictive interaction between the squared order parameter and the strain caused by the acoustic waves\cite{Rehwald1973}. The detailed discussion of the temperature dependence of the elastic constants is published elsewhere \cite{Ko2013,Ko2014determination}.
It is instructive to note that at high temperatures the crystal appears to be highly isotropic with regard to the initial slope of transverse phonons: the values of speed of sound along [1 0 0] and [1 1 0] are nearly identical. This means that the TA phonons in $\Gamma$-X and $\Gamma$-M directions have the same dispersion close to the $\Gamma$-point. This makes the huge anisotropy in TA dispersion that emerges at higher $q$-values especially remarkable.

\newpage

\section{Soft modes and central peaks}\label{CP}
As seen from the results presented in the previous section the observed TO mode resonance is practically temperature independent and cannot be considered as the origin of the growth of the dielectric permittivity. 
Instead of the mode softening we see the growth of the central peak (CP). Ostapchuk et al. \cite{Ostapchuk2001} observed in the IR absorption spectra a similar feature - an independent critical 'central mode' responsible for the dielectric anomaly. However it is well established  that the critical growth of the CP is not necessarily related to an additional degree of freedom critically slowing down but can result from the coupling of the soft mode with intrinsically temperature independent degree of freedom.
Such approach was developed during last 40 years and is widely discussed in the literature \cite{Buixaderas2004,Riste1971,Bruce1981,Axe1973}.

Following Ref. \onlinecite{Axe1973} we write the expression for the spectral correlation function $A_{\lambda}$ in quasiharmonic approximation, where $\lambda$ is a composite index including the reduced wavevector and mode ``number'' :

\begin{equation}
A_{\lambda}(\omega )=\frac{1}{\pi } Im \left[ \Omega_{\lambda}^2 - \omega^2 +\Pi_{\lambda}^{an} (\omega) \right]^{-1}
\end{equation}
Here $\Omega_{\lambda}$ stands for undisturbed harmonic frequency and $\Pi_{\lambda}^{an} (\omega) $ stands for self-energy function.
\begin{equation}
\Pi_{\lambda}^{an} (\omega) =\Delta_{\lambda}^0+i\omega\Gamma_{\lambda}^0,
\end{equation}
$\Delta_{\lambda}^0$ \ and $\omega\Gamma_{\lambda}^0$ are frequency independent constants. This formula corresponds to the classical damped harmonic oscillator with a characteristic "quasiharmonic" frequency:
\begin{equation}
\label{renormalization}
\omega_{\infty}^2=\Omega^2 + \Delta^0,
\end{equation}
hereafter the index $\lambda$ is omitted.
The appearance of the central peak can be accounted for by introducing the frequency dependence of the parameters $\Delta$\ and $\Gamma$, which is presented in the form:
\begin{eqnarray}
\Delta(\omega) & = \Delta^0 &-\delta^2\left[\frac{\gamma^2}{\gamma^2+\omega^2}\right]\\
\Gamma(\omega) & = \Gamma^0 &+\delta^2\left[\frac{\gamma}{\gamma^2+\omega^2}\right]
\end{eqnarray}
The introduction of such frequency dependences results in a 2-component response.
There is a double-peak damped harmonic oscillator component:
\begin{equation}
S_{\mathrm{DHO}}(\omega)=\frac{\Gamma^0}{(\omega_{\infty}^2-\omega^2)^2+(\omega\Gamma^0)^2}
\end{equation}
and a central peak component
\begin{equation}
S_{\mathrm{CP}}(\omega)=\left(\frac{\delta^2}{\omega_0^2\omega_{\infty}^2}\right)\frac{\gamma '}{\omega^2+\gamma '^2}.
\end{equation}
Here
\begin{equation}
\label{renormalization1}
\omega_0^2=\omega_{\infty}^2-\delta^2
\end{equation} 
and $\gamma '=\gamma(\omega_0/\omega_{\infty})^2$.

In this framework the quasiharmonic frequency $\omega_{\infty}$ of the TO mode can be substantially renormalized (see Eq. (\ref{renormalization})) with respect to the effective soft mode frequency $\omega_0$, whose temperature dependence defines the critical process.
Due to that, in the experiment, one may observe nearly temperature-independent phonon resonances that are accompanied by a strongly temperature dependent central peak manifesting the critical lattice softening. 

\begin{figure}
\includegraphics [width=.8\columnwidth,clip=true, trim=0mm 0mm 0mm 0mm] {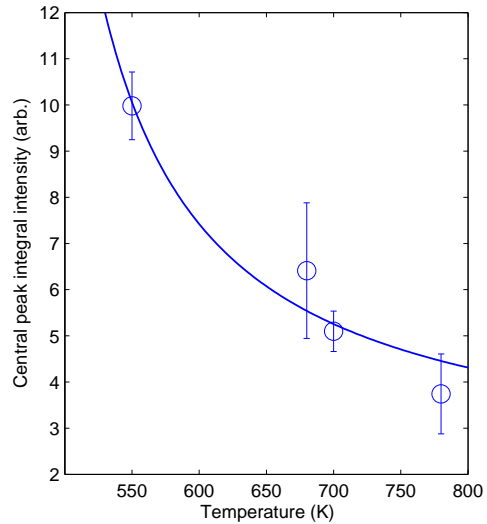}
\caption{\label{fig_cp}
Temperature dependence of the central peak integral intensity. The dependence corresponds to the the momentum transfer $\vec{Q}$=(2.15~2.85~0) (reduced scattered wavevector $\vec{q}$=(0.15~-0.15~0)). Error bars correspond to 95\% confidence interval. Fit to Eq.~(\ref{CPcrit}) is shown as solid line.
}
\end{figure}

Returning to the dielectric anomaly and the $\Gamma$-point soft mode we find that the temperature dependence of the central peak intensity is consistent with the assumed critical decrease of the $\omega_0$ frequency of the ferroelectric soft mode. Figure \ref{fig_cp} shows the fit of the experimental data to the formula \citep{Bruce1981}
\begin{equation}\label{CPcrit}
I^{\textrm{crit}}(q,T)=\frac{T}{T-T_0 + gq^2}.
\end{equation}

This formula describes the expected temperature dependence of the central peak intensity due to the condensation of the soft mode with critical temperature $T_0=485$~K \cite{Shirane1951}. Parameter $g$ is connected to the correlation length as $r_c=\sqrt{g\over{T-T_c}}$. 
Thus the central peak is indeed a signature of the PZO's ferroelectric lattice instability that is reflected by the dielectric anomaly. 
Rattling motion of the Pb$^{2+}$\ ions in the multi-well potential can be tentatively considered as the microscopic realization of the slow relaxing degree of freedom. 
This rattling is not the origin of the phase transition but strongly renormalizes the critical dynamics.

\section{R-point}
\begin{figure}
\includegraphics [width=\columnwidth,clip=true, trim=0mm 0mm 0mm 0mm] {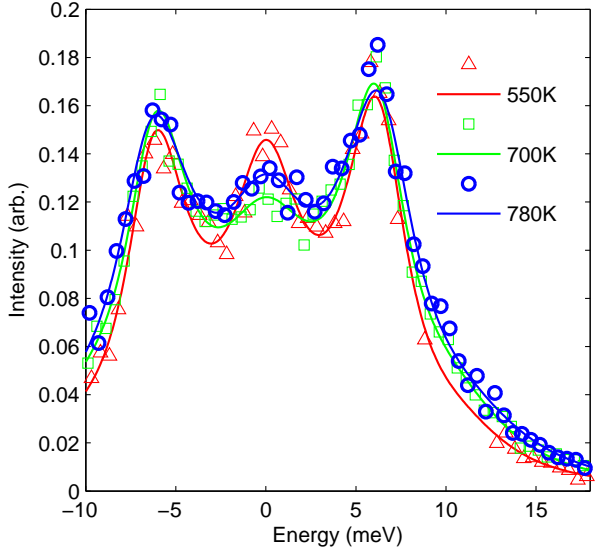}
\caption{\label{fig_r_spectra}
Experimentally measured IXS spectra at the R-point ($\vec{Q}$=(3.5 0.5 0.5)) at several temperatures.
 $\vec{Q}$  is measured in the units of the reciprocal cubic lattice constant $a^*=2\pi/a$.}
\end{figure}

Figure \ref{fig_r_spectra} shows the evolution of the R-point IXS spectra at different temperatures. These spectra consist of the normal phonon resonances and the central peak. We can see no temperature dependence of the phonon peaks, while the central peak demonstrate weak increase on approaching the transition temperature.

It is important to estimate whether the observed growth of the CP can be considered as an evidence of the instability relevant to the observed phase transition. 
In compounds containing heavy atoms it may be difficult \cite{Corker1997} to reliably analyze the oxygen-involved zone-boundary distortions by X-ray scattering due to small atomic scattering factor of oxygen. 
However, as we show below, in the case of inelastic scattering this is partially compensated by the small mass of the oxygen atoms, providing reasonable ratio of the scattering intensities due to modes dominated by the motion of heavy ions and oxygen-related modes.

In the harmonic approximation the dynamical structure factor for one-phonon scattering can be presented as \cite{Krisch2007}
\begin{equation}
\label{DynamicalSF}
S(\vec{Q},\omega) = \frac{1}{2}\sum_j \langle n(\omega) + \frac{1}{2} \pm \frac{1}{2} \rangle
\frac{1}{\omega_j(q)}F_{jn}(\vec{Q})\delta(\omega \pm \omega_j(\vec{q})),
\end{equation}
where $\omega_j$ is the frequency of the $j$-th mode, $n(\omega)$ - the Bose factor and $F_{in}(\vec{Q})$ - the inelastic structure factor. The latter is given by a sum over the atoms in the unit cell
\begin{equation}
\label{InelasticSF}
F_{jn}(\vec{Q}) =
\left| \sum_a \frac{f_a(\vec{Q})}{\sqrt{M_a}}
[\vec{e}_a^j(\vec{q})\cdot \vec{Q}]
\textrm{exp}(i\vec{Q} \cdot \vec{r}_a) \text{exp}(-w_a)\right|^2.
\end{equation}

Here $M_a$ represent masses of the ions, $f_a(\vec{Q})$ - the atomic scattering form factors, $\vec{e}_a^j(\vec{q})$ - the mode eigenvector, $\vec{r}_a$ - positions of the atoms in the unit cell and $\textrm{exp}(-w_a)$ - the Debye-Waller factors, the suffix $a$ enumerates the atoms in the unit cell.
To obtain an estimate for possible relative intensities of scattering due to vibrations of heavy ions (soft mode) and light ions (oxygen mode) we roughly evaluate the contributions to the inelastic structure factor (\ref{InelasticSF}) due to lead and oxygen atoms.
For simplicity we neglect the difference in the Debye-Waller factors of the elements and use the approximation $f_a(Q) \approx f_a(0)= Z_a$ (where $Z_a$ is the number of the element in the periodic table), which is reasonable for not too large $Q$.
We obtain in this approximation the estimates for the largest possible contributions of these elements to the inelastic structure factors.
In this case, the order of magnitude of the leading term in (\ref{InelasticSF}) for the lead for $\vec{Q}= (3\ 2\ 0)$  will be $(\frac{Z_\text{Pb}}{\sqrt{M_\text{Pb}}}*Q)^2= 422.28$ (with $Z_\text{Pb}= 82$ and $M_\text{Pb}= 207$) while for the oxygen at $\vec{Q}= (3.5\ 0.5\ 0.5)$ $(\frac{Z_\text{O}}{\sqrt{M_\text{O}}}*Q)^2= 51$ (with $Z_\text{O}= 8$ and $M_\text{O}= 16$).
From these estimates we can conclude that the scattering intensities of a lead containing mode and of a purely oxygen one, once these are of comparable frequencies, should differ less than one order of magnitude.

\begin{figure}
\includegraphics [width=\columnwidth,clip=true, trim=0mm 0mm 0mm 0mm] {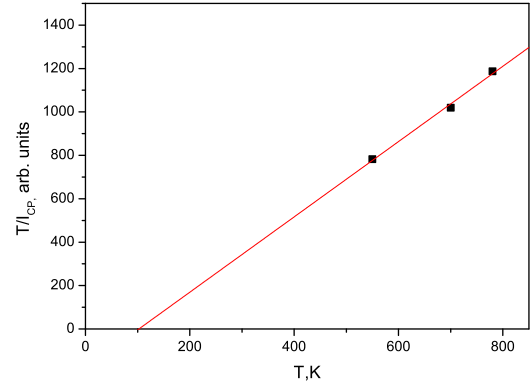}
\caption{\label{fig_r_extrapolation}
$T/I_{CP}(T)$ dependence, where $I_{CP}$ - integrated intensity of the central line at R-point ($\vec{Q}$=(3.5 0.5 0.5)).
$\vec{Q}$  is measured in the units of the reciprocal cubic lattice constant $a^*=2\pi/a$.}
\end{figure}

Indeed our measurements at the mentioned R-point, presented in  Fig. \ref{fig_r_spectra}, give distinct spectra containing phonon resonances which we can tentatively identify with the overdamped R$_{25}$ mode, and showing the central peak. In contrast with the temperature independent phonon resonances, the central peak demonstrates a traceable evolution with temperature.
To quantify the evolution of the corresponding mode frequency we employ the fact that at high temperatures the temperature-normalized integral central peak intensity $I_\text{CP}/T$ is proportional to the generalized static susceptibility \citep{Bruce1981}. The temperature evolution of $T/I_\text{CP}$ is presented in Fig. \ref{fig_r_extrapolation}.
Apparently, the softening of the R$_{25}$ mode does not correspond to any critical process with characteristic temperature close to the transition temperature (503 K). The extrapolation of the temperature trend of $T/I_\text{CP}$ gives the critical temperature about 100 K.

We can also provide an estimate of the absolute value for the R-point ``unperturbed'' phonon frequency $\omega_0$. It is the smallness of this value, which gives us the information about the potential instability of the crystal with respect to the relevant phase transformation.
In view of the presence of the central peak, this frequency differs from that of the phonon resonance.
We assume that the normal phonon resonances and the central peak at R-point are both related to the same mode with 'unperturbed' frequency $\omega_{0}$. 
The latter is connected with the frequency $\omega_{\infty}$, at which the phonon resonances are observed, by Eq. (\ref{renormalization1})
where the parameter $\delta$  can be determined through the relationship \cite{Axe1973}
\begin{equation}
\label{intensity_ratio}
\frac{I_\text{CP}}{I_\text{Total}}=\frac{\delta^2}{\omega_{\infty}^2}.
\end{equation}
Here $I_\text{CP}$ and $I_\text{Total}$ are integral intensities for the central peak and for the whole spectrum, respectively.
Relationships (\ref{renormalization1}) and (\ref{intensity_ratio}) imply $\omega_0^2=\omega_{\infty}^2(1-I_\text{CP}/I_\text{Total})$.
At 780~K from our experimental data the ratio $\frac{I_\text{CP}}{I_\text{Total}}=0.72\pm0.03$ and the energy of phonon resonance $\omega_{\infty}=6.6\pm0.3$~meV. This gives the ``unperturbed'' mode frequency $\omega_{0}=3.5\pm0.35$~meV. At 550~K the result is the same within the error bars. Apparently this frequency is sufficiently high in comparison with the frequency of the TA mode close to the $\Sigma$-point thus making it unlikely to be a relevant instability with regard to the AFE phase transition at $T=503$ K.


\section{Modulation instability}
In the previous chapters we identified the following: 
(1) There is no temperature-dependent TO phonon resonance, and the soft TO mode -- that is responsible for the dielectric anomaly -- is manifested in IXS as a temperature-dependent central peak. 
(2) There is a temperature-dependent TA mode, whose temperature dependence is much sharper than could be suggested from the temperature dependence of elastic constants.
The next step is to figure out the relationship between these two observations. The idea is to connect the unperturbed frequency $\omega_0$ of the soft TO mode with the frequency of the TA mode in the same fashion as it was done in other perovskites. Indeed the TA-TO mode coupling was identified as the source of ``pushing down'' of the TA mode in KTaO$_3$ \cite{Farhi2000} and other crystals with ferroelectric soft modes. Since the piezoelectric coupling is forbidden by symmetry in the paraelectric phase, the interaction between TA and TO modes is described not by bilinear coupling of polarization and deformation but by gradient terms involving these quantities in the free energy expansion. The latter reads

\begin{widetext}
\begin{equation}
F= \frac{\alpha}{2} P^2 + \frac{c_{ijkl}}{2} u_{ij}u_{kl}+
\frac{\\g_{ijkl}}{2}\frac{\partial P_{i}}{\partial x_j}\frac{\partial P_{k}}{\partial x_l} 
- \frac{\\f_{ijkl}}{2} \left(P_k\frac{\partial u_{ij}}{\partial x_l} -
u_{ij}\frac{\partial P_k}{\partial x_l}\right)
\label{flex}
\end{equation}
\end{widetext}
where $\alpha= A(T-T_0)$, $P_{i}$ - components of the polarization vector and $u_{ij}$ - components of the elastic strain tensor. Hereafter the Einstein summation convention is adopted.
The flexoelectric tensor $f_{ijkl}$ describes the interaction between polarization and gradient of deformation and the interaction between gradient of polarization and deformation. As it is shown later the flexoelectric coupling works only at finite wavevectors $q$, leaving the initial slope of the elastic waves dispersion intact. The effect of this coupling on the finite-$q$ phonon spectrum is a ``repulsion'' between the frequencies of the TA and TO modes. 

The presence of the flexoelectric term in the free energy expansion (\ref{flex}) makes the system potentially unstable with respect to spatial modulations of polarization and strain.
The criterion for the appearance of such instability, formulated in terms of phonon eigenstates, was proposed by Axe \textit{et al} \cite{Axe1970}.
These authors also supposed that ``the materials exhibiting such instability existed or would be found''.
For cubic (perovskite) materials the modulation instability criterion can be readily rewritten in terms of the $c$, $g$, and $f$ tensors.
For the modulations along the [1 1 0] direction there are possible two such formulations depending on the polarization vector \cite{Yudin2014}. For the in-plane polarization (along [1 -1 0]) the criterion reads 
\begin{equation} \label{eq_theta1}
\Theta_1=(f_{11}-f_{12})/\sqrt{(c_{11}-c_{12})(g_{11}-g_{12})}; \Theta_1 < 1, 
\end{equation}
while for the out-of-plane polarization (along [0 0 1]) it is
\begin{equation} \label{eq_theta2} 
\Theta_2=f_{44}/\sqrt{c_{44}g_{44}}; \Theta_2 < 1.
\end{equation}
For typical perovskite ferroelectrics BaTiO$_3$ and SrTiO$_3$, using experimental data \cite{Eggenhoffner1982,Vaks1973,Tagantsev2001}, these criteria can be substantialized \cite{Yudin2014} as $|f_{44}|< 3.3\,\textrm{V}$, $|f_{11}-f_{12}|<7\,\textrm{V}$ and $|f_{44}|< 2.4\,\textrm{V}$, $|f_{11}-f_{12}|<10\,\textrm{V}$, respectively.
According to order-of-magnitude estimates \cite{Kogan1964,Zubko2013,Yudin2014} and \textit{ab initio} calculations in perovskites \cite{Ponomareva2012}, the values of the components of the $f$ tensor in perovskites are expected to be about $1-10$ V.
Thus, we see that ferroelectric perovskites are not far from the modulation instability.

The fact that the apparent ``pushing down'', or renormalizing of the TA mode in PZO is much more pronounced than in other studied systems suggests that it could be the prime candidate in the search for materials where the modulation instability ``would be found''. Triggering of such instability would happen if the TA branch would touch (in the case of the second order transition) or ``almost touch'' (in the case of the first order transition) the zero energy level at some $q$-point along the $\Gamma-\Sigma$ direction. In this case the newly formed modulated phase should be characterized by the same or similar ionic displacements as would have a frozen renormalized TA phonon in the corresponding $q$-point. And indeed, the AFE phase of PZO has the pattern of the lead displacements consistent with the scenario in which it is formed as the result of the TA mode condensation at $q = q_\Sigma$.
This pattern corresponds to the so-called  $\Sigma_3$ normal mode of the cubic structure \cite{Fujishita1984,Cowley1964}.
For the wave vector $\vec{q} =\vec{q}_\Sigma$ one can present the lead displacements in such mode in the form:
\begin{equation}
\vec{u}_{Pb} \propto
\begin{pmatrix}
-1 &\\
  1 &\\
  0 &
\end{pmatrix}
\cos \left[\frac{\pi}{2a}(x+y)+\varphi \right]
\label{Pb}
\end{equation}
where $x$ and $y$ are the lead ions Cartesian coordinates in the cubic reference frame.
One can readily check that the pattern of the lead displacements in the low-temperature phase is reproduced by Eq.~(\ref{Pb}) for $\varphi =\pi/4, 3\pi/4,5\pi/4,7\pi/4$. These four values of $\varphi$ correspond to four translational domain states of one orientational domain state specified by the orientation of the modulation wave vector.

In fact the transition is of the first order, the modulation vector is $\vec{q} =\vec{q}_\Sigma$ and the TA branch is softening as a whole, without any point on it being special. Consequently one needs to find a mechanism by which the modulation with this particular commensurate wavevector occurs instead of modulations with the arbitrarily close incommensurate variants. In systems with real incommensurate phases it often happens that the generally incommensurate modulation is pinned down to the particular commensurate wavevector. The terms in the free energy expansion responsible for this effect are called Umklapp terms \cite{Bruce1981,Tagantsev2013}. In PbZrO$_3$ we assume the Umklapp terms to be strong enough to drive the phase transition directly to the commensurate AFE state, skipping the incommensurate phase.

\section{Diffuse scattering}
The outlined in the previous section scenario of the flexoelectricity-driven change of the TA phonon group velocity on going from small to large $q$ along [1 1 0] direction can in principle be tested by examining the frequencies and eigenvectors of the low-energy phonon modes \cite{Farhi2000}. In PZO the IXS technique does not allow to directly analyze the dispersion of  the ferroelectric mode. Instead, this mode is manifested in the temperature-dependent central peak. Another complementary technique that nevertheless provides the means for checking the proposed  concept is the diffuse scattering. By diffuse scattering (DS) we mean here the energy-integrated diffuse scattering as opposed to the energy-resolved DS that was numerously reported in studies of relaxors (see Refs. \onlinecite{Gehring2009,Cowley2013} for reviews). With energy-integrated approach the intensity of the DS comprises the contributions from all the phonon modes. The observed central peak can be included in the consideration following the ideas outlined in the section \ref{CP}. In this section we present the model and experimental data on the diffuse scattering in PZO. The calculation and analysis of lattice dynamics is simplified assuming the limit of long wavelengths and neglecting high-energy optic modes that are irrelevant here \cite{Vaks1973,Farhi2000}. The resulting simplified Hamiltonian takes into account only 5 modes: 3 acoustic modes (2TA+LA) and 2 lowest-energy transverse optic modes (2TO). It reads \cite{Vaks1973,Farhi2000}:

\begin{widetext}
\begin{eqnarray}\label{5mode}
{\mathcal H}^{(5)}= 
\frac{1}{2}\sum_q 
\left[ \dot{\mathbf u}_{-q}\dot{\mathbf u}_{q} +
{\mathbf u}_{-q}\hat{A}(q){\mathbf u}_{q} + 
\dot{\mathbf x}_{-q}\dot{\mathbf x}_{q} +
\omega_0^2 {\mathbf x}_{-q} {\mathbf x}_{q} + 
{\mathbf x}_{-q}\hat{S}(q){\mathbf x}_{q} +
{2\mathbf u}_{-q}\hat{V}(q){\mathbf x}_{q} \right]
\end{eqnarray}
\end{widetext}
where $u_1,u_2,u_3$  and  $x_1,x_2$ are the normal coordinates for the 2TA+LA and 2TO modes, in the reference frame ($X'Y'Z'$) with $Z'$-axis parallel to the reduced wavevector  $\textbf{q}$, respectively.
The tensors $\hat{A}$, $\hat{S}$ and $\hat{V}$ describe the contribution of the short-range interactions and can be written as:
$
\hat{A} =  q^2(A_l \hat{g}^l  +  A_t \hat{g}^t  +  A_a \hat{g}^a),\
\hat{S}  =  q^2(S_t \hat{g}^t  +  S_a \hat{g}^a),\
\hat{V}  =  q^2( V_t \hat{g}^t  + V_a \hat{g}^a),  	
$
where
$
\hat{g}^{l}_{\alpha\beta}  =  n_{\alpha}n_{\beta},\
\hat{g}^{t}_{\alpha\beta}  =  \delta_{\alpha\beta} - n_{\alpha}n_{\beta},\
\hat{g}^{a}_{\alpha\beta}  =  \gamma_{\alpha\beta\gamma\delta}n_{\gamma}n_{\delta}.
$
In these equations ${\mathbf n}={\mathbf q}/q$, $\delta_{\alpha\beta}$ is the Kronecker delta, and $\gamma_{\alpha\beta\gamma\delta}$ is the tensor invariant with respect to the symmetry operations of the cubic point groups, which, in the cubic reference frame, is defined  as $\gamma_{\alpha\beta\gamma\delta}=1$\ for $\alpha=\beta=\gamma=\delta$\ and $\gamma_{\alpha\beta\gamma\delta}=0$\ otherwise.
$\omega_0$ is the energy of the $\Gamma$-point soft mode.
The parameters of the matrices $\hat{A}$, $\hat{S}$, and $\hat{V}$\ can be related to the coefficients of the free energy expansion (\ref{flex}):
$A_l = (c_{12} + 2c_{44})/\rho$,
$A_t = c_{44}/\rho$,
$A_a = (c_{11} - c_{12} - 2c_{44})/\rho$,
$S_t = g_{44}/\mu$,
$S_a = (g_{11} - g_{12} - 2g_{44})/\mu$
$V_t = f_{44}/\sqrt{\mu\rho}$,
$V_a = (f_{11} - f_{12} - 2f_{44})/\sqrt{\mu\rho}$
where $\rho$ is the density and $\mu$ is the coefficient relating $\omega_0$ with the inverse electric susceptibility $\alpha$ in the cubic phase: $\alpha=\mu\omega_0^2$.
The relationships for $V_t$ and $V_a$ are approximate since they ignore the contribution of the dynamic flexoelectricity \cite{Tagantsev1986}.

The standard solution to the dynamical problem with Hamiltonian (\ref{5mode}) gives the frequencies and eigenvectors of the renormalized lattice modes. The latter can be presented in the reference frame ($X'Y'Z'$) as the columns of the matrix

\begin{equation} \label{eig}
D =
\begin{pmatrix}

 v_{TOX'}^{(1)} & v_{TOX'}^{(2)} &  v_{TOX'}^{(3)} & v_{TOX'}^{(4)} & v_{TOX'}^{(5)} \\
 v_{TOY'}^{(1)} & v_{TOY'}^{(2)} &  v_{TOY'}^{(3)} & v_{TOY'}^{(4)} & v_{TOY'}^{(5)} \\
 v_{TAX'}^{(1)} & v_{TAX'}^{(2)} &  v_{TAX'}^{(3)} & v_{TAX'}^{(4)} & v_{TAX'}^{(5)} \\
 v_{TAY'}^{(1)} & v_{TAY'}^{(2)} &  v_{TAY'}^{(3)} & v_{TAY'}^{(4)} & v_{TAY'}^{(5)} \\
 v_{LAZ'}^{(1)} & v_{LAZ'}^{(2)} &  v_{LAZ'}^{(3)} & v_{LAZ'}^{(4)} & v_{LAZ'}^{(5)} \end{pmatrix}. 
\end{equation}

The values $v_{S\alpha}^{(N)}$ describe the contributions of the unperturbed modes of the type $S$ polarized along the axis $\alpha$\ in the ($X'Y'Z'$)\ frame to the mode $N$\ in the coupled system. In the case of no mode coupling all $v_{S\alpha}^N$\ are either 1 or 0.
Eigenvectors in the original cubic crystallographic reference frame ($XYZ$) are restored by a known transformation, described by matrix ${\mathbf M}$\ in Refs. \onlinecite{Vaks1973,Farhi2000}.

In the frame ($XYZ$), the eigenvectors of (\ref{5mode}) will have 6 components ($w_1^{(i,1)}$, $w_2^{(i,1)}$, $w_3^{(i,1)}$, $w_1^{(i,2)}$, $w_2^{(i,2)}$, $w_3^{(i,2)}$) where $i=1..5$ enumerates the eigenmodes and the upper scripts "1" and "2"  specify the Cartesian components of the  contributions from the unperturbed optic ("1") and acoustic ("2") modes. The intensity of integral diffuse scattering due to the considered modes is given by
\begin{equation}\label{diffsc}
I({\mathbf Q}) \propto T\sum_{i=1}^5\sum_{l=1}^2\frac{1}{\omega^2_i({\mathbf q})}|{\mathbf Q \cdot \textbf{w}^{(i,l)}(q)}|^2.
\end{equation}
Here $\omega_i({\mathbf q})$\  are the frequencies  of the renormalized phonon modes.
The presented expression is similar to the usual one \cite{Bruce1981}, but there are no phase factors $\exp i{\mathbf Q \cdot \textbf{r}_j}$ while the $1/\sqrt{m_j}$ multipliers are to be included in the eigenvectors of unperturbed modes.
Here $\textbf{r}_j$ and $m_j$ are the radius-vector and the mass of the $j^{th}$ atom in the unit cell, respectively.
We expect this formula to give the correct description of the shape of the DS around any reciprocal lattice point, while it cannot be used for the comparison of the intensity around different Bragg peaks.

\begin{figure}[ht]
\includegraphics [width=1\columnwidth,clip=true, trim=0mm 0mm 0mm 0mm] {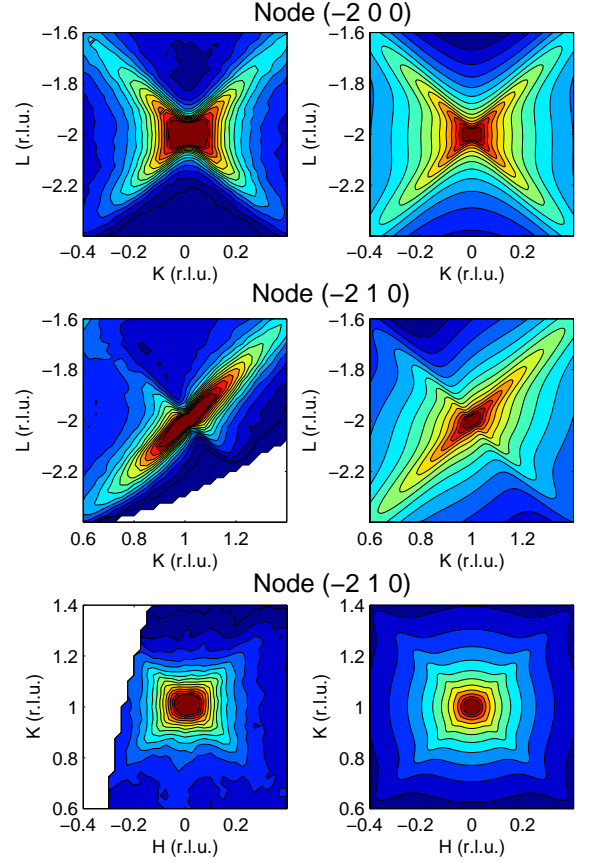}
\caption{\label{fig_diffuse}
Experimental (left column) and modeled (right column) diffuse scattering distributions for nodes (-2 0 0) and (-2 1 0). Temperature $T=550$ K.
}
\end{figure}

We checked the consistence of this model with the experimentally observed intensities. The latter are shown in Fig. \ref{fig_diffuse} in the left column. One can see a substantially anisotropic distributions similar in shape to ones in relaxor ferroelectrics. The shining rods along [1 1 0] correspond to the $q$-vectors where the in-plane TA mode is the softest. The results of our modeling are shown in the right column, one may note a good qualitative agreement.

During modeling we had to deal with eight parameters: $A_l$, $A_t$, $A_a$, $S_t$, $S_a$, $V_t$, $V_a$ and $\omega_0$. We simulated the 2D intensity map for 550 K.
We used the value of $\omega_0 = 1.25$ meV at this temperature from the results of infrared measurements of Ostapchuk \textit{et al}. \cite{Ostapchuk2001}.
The parameters controlling the dispersion of non-coupled acoustic branches  can be recalculated on the basis of elastic constants c$_{11}$=194 GPa, c$_{12}$=61 GPa, c$_{44}$= 71 GPa determined from the data of our Brillouin scattering experiment, density $\rho= 8$ g/cm$^3$, and the cubic lattice constant $a = 0.416$ nm \cite{Whatmore1979} of PZO:
$A_l =$ 2512 meV$^2$/rlu$^2$
$A_t = $ 879 meV$^2$/rlu$^2$
$A_a = $ -111 meV$^2$/rlu$^2$.
The parameters controlling the dispersion of the  TO mode ($S_t$ and $S_a$) cannot be directly determined by inelastic scattering because this mode is overdamped in PZO.
In modeling we assume that the isotropic part of the TO dispersion is close to one in lead titanate \cite{Kempa2006} and use the value $S_t$=4330 meV$^2$/rlu$^2$. We restricted our consideration to isotropic coupling $V_a = 0$. The values of unperturbed TO anisotropy $S_a = -7100$ meV$^2$/rlu$^2$ and the isotropic coupling constant $V_t = 750$ meV$^2$/rlu$^2$ were determined by nonlinear least-squares fit of diffuse scattering distribution. 

The magnitudes of the latter two values are similar to ones in other perovskites. In fact, the magnitude of coupling $V_t$ is about two times smaller than for PbTiO$_3$ \cite{Vaks1973}.
The absolute value of the anisotropy parameter $S_a$ is slightly smaller than in KTaO$_3$ and  2-3 times smaller than in BaTiO$_3$ \cite{Vaks1973}.
What makes the qualitative difference is the sign of $S_a$.
Due to a positive $S_a$ in  KTaO$_3$ and BaTiO$_3$, the frequency for the out-of-plane polarized TA phonons is minimal for the $\{001\}$ planes, resulting in the observed shining diffuse planes.
In PbZrO$_3$ the parameter $S_a$ is negative and  the frequency of the in-plane polarized TA phonons, propagating along $\langle 110\rangle$ directions is minimal, as a result the shining diffuse rods are observed.
We have to admit that the presented values of parameters are not unique especially because the parameters $S_a$ and $V_a$ are correlated. The most important feature of the model is a mode coupling constant. As it is mentioned above the obtained value for it is quite typical for the ferroelectric (real or incipient) perovskites. 

One can check that the spectral parameters used for simulation of the 2D intensity maps corroborate with the statement that PZO is not far from the incommensurate instability. The parameter $\Theta_1$ (Eq. (\ref{eq_theta1})) can be readily expressed in terms of these parameters to find $\Theta_1=(2V_t+V_a)/\sqrt{(2A_t+A_a)(2S_t+S_a)}=0.9$.
Thus $\Theta_1$ is close to its critical value of 1.
However, in view of neglecting the contribution of the dynamic flexoelectricity when linking $f's$ with $V's$, the value of $\Theta_1$ obtained should be taken with precaution.


\section{Antiferroelectricity}
We have shown that the anti-polar lead displacements giving the quadrupling of the unit cell can be viewed as a result of the softening of a zone-center ferroelectric mode.  At the same time, as it should be in antiferroelectric, the ferroelectric instability does not lead to the formation of a stable ferroelectric state in PZO.  A natural question is of why it happens. What is the mechanism for the suppression of the ferroelectric instability? This can be readily  explained in terms of a two-mode model with competing interactions \citep{Balashova1993,Tagantsev2013}. For PZO, these two modes are the zone-center ferroelectric  polar mode with polarization $P$ as the order parameter and the $\Sigma_3$ mode associated with the complex order parameter describing six orientational domain states and four translational domain states in the low-symmetry phase \cite{Tagantsev2013}. For the orientational domain state defined by the modulation wavevector $\vec{q­}_\Sigma = (1/4\ 1/4\ 1/4)$ and translational domain state defined by a phase  $\phi$ taking the values $\pi/4$, $3\pi/4$, $5\pi/4$ and $7\pi/4$ this order parameter is expressed as \cite{Tagantsev2013}
\begin{equation} \label{XOP}
\xi = \rho \exp(i \vec{q}_\Sigma \vec{r} + i \phi),
\end{equation}
 where $\rho$ is the modulus of the order parameter. Taking into account  the symmetry of the problem, the free energy describing the aforementioned completing interaction reads: 

\begin{eqnarray}\label{F_AN}
F(P,\xi)=\frac{1}{2}A(T-T_0)P^2 + \\
 \frac{1}{2}\delta_\text{P1} (P_1^2+P_2^2)\rho^2 +\frac{1}{2}\delta_\text{P3} P_3^2 \rho^2 \nonumber \\ + F_\text{A}(\xi) \nonumber 
\end{eqnarray}
The term $F_\text{A}(\xi)$ describes the free energy associated with the structural order parameter and accommodates the terms describing Umklapp interaction that triggers the transition and the constants $\delta_\text{P1}$ and $\delta_\text{P3}$ define the repulsive biquadratic coupling between the polarization and the structural order parameter. The equation of state for polarization $\partial F/\partial P_i=E_i$ ($E_i$ is a component of electric field vector), for the high-temperature phase (with $\xi=0$), yields the Curie-Weiss law $\chi \propto1/(T-T_0)$ for the dielectric susceptibility defined as $\chi=dP/dE$. For the low-temperature phase where the order parameter of the transition, $\xi$, acquires a spontaneous value described by (\ref{XOP}), the  susceptibility is described by a diagonal tensor with elements
\begin{equation}\label{eq_chi1}
\chi_{ii}=\frac{1}{A(T-T_0)+ \delta_\text{Pi} \rho_0^2},
\end{equation}

where $\rho_0$ is the modulus of the spontaneous value of the structural order parameter $\xi_0 = \rho_0 \exp(i \vec{q}_\Sigma \vec{r} + i \phi)$. Here the subscripts in $\chi_{ii}$ do not mean summation.
Equations (\ref{eq_chi1}) correspond to the aniferroelectric-type anomaly if the denominators $A(T-T_0)+ \delta_\text{Pi} \rho_0^2$ increase on cooling.
This is possible if the increase of $\rho_0$ with lowering temperature dominates the behavior of these terms.
Such a condition can be assured by a large enough coupling constants $\delta_\text{P1}$ and $\delta_\text{P2}$.

Thus, we see that the ferroelectric instability does not end up with a ferroelectric phase for it is suppressed by the developed structural order parameter. This instability, being the driving force for the appearance of  $\xi_0$, is suppressed by it at $T < T_\text{A}$. It is shown in Ref. \onlinecite{Balashova1993} that the application of the electric field effectively shifts the transition temperature down which allows to interpret the double-hysteresis loops as a result of a field-induced transition between AFE and externally polarized parent phases.

The two-instability model also allows an antiferroelectric to ferroelectric phase transition given a small chemical modification or the application of hydrostatic pressure.
Indeed, in view of the small difference between the temperatures $T_0$ and $T_\text{A}$, which is the key element of the model, the aforementioned factors may swap the relative positions of these temperatures.
In this case the ferroelectric phase transition will take place first on cooling.
Such a scenario corroborates  with the appearance of ferroelectricity once PZO is slightly doped with Ti \cite{Jaffe1971}.

The two-instability model also naturally predicts the positivity of development of ferroelectricity in anti-ferroelectric domain walls \cite{Wei2014}. Indeed, if in such wall the structural order parameter passes through zero, as is clear from Eq. (\ref{F_AN}), its suppressing effect on the ferroelectric instabilty  inside the wall is strongly reduced, making the development of the ferroelectricity favorable. This possibility was recently confirmed by Wei $et\ al.$ \cite{Wei2014} who documented  the development of ferroelectricity in AFE domain walls in PZO, using transmission electron microscopy data.

\section{Octahedra rotations} 

We complete the picture of phase transition in PZO by addressing the mechanism by which the additional oxygen octahedra tilts do develop below $T_A$. We can describe these rotations as being triggered by the Holakovsky mechanism \cite{Holakovsky1973}, which was recently identified in perovskite ferroelectrics \cite{Kornev2009}.
Similarly to the above treatment of the suppressed ferroelectricity we consider the expansion of the free energy in terms of the order parameter (\ref{XOP}) describing lead displacements for a given orientational domain state and another order parameter - a real pseudovector $\vec{\Phi}$ describing the oxygen-octahedron rotations \cite{Tagantsev2013}
\begin{eqnarray}
F_\Phi=\frac{1}{2}\alpha_\Phi\Phi^2 + \frac{1}{2}\delta_{\Phi1} (\Phi_1^2+\Phi_2^2)\rho^2 \\ \nonumber +\frac{1}{2}\delta_{\Phi3} \Phi_3^2 \rho^2+F_{\Sigma} +  \textrm{high-order terms in}~ \Phi.
\label{F_Fi}
\end{eqnarray}
where $F_{\Sigma}$ is a function of the $\Sigma$-point order parameter only.
If at least one of the coupling constants $\delta_{\Phi1}$ and $\delta_{\Phi3}$ is negative, the appearance of the spontaneous order parameter of the transition, $\xi_0$, may trigger that of the order parameter $\Phi$.
This happens if at the transition point either $\alpha_\Phi + \delta_{\Phi1}\rho_0^2$  or $\alpha_\Phi + \delta_{\Phi3}\rho_0^2$ is negative.
In the case of PZ $\Phi_1= \Phi_2\neq0$ while $\Phi_3= 0$ \cite{Whatmore1979}.
The considered phenomenological framework describes this situation if $\delta_{\Phi1}<0$ and at the transition $\alpha_\Phi + \delta_{\Phi1}\rho_0^2<0.$
Thus, in our scenario, the oxygen-octahedron rotations in PZO are induced by the anti-polar lead displacements via an attractive biquadratic mode coupling.


\section{Conclusion}
Antiferroelectricity is a rather complex phenomenon and the assessment of its microscopic mechanism is challenging. Here we have drawn a picture consisting from the experimental facts about the lattice dynamics and the basic ideas on the interpretation of these facts in PbZrO$_3$. The experimental data show that the pronounced slowing down of the lattice dynamics on approaching the phase transition is due to the softening TO mode which induces the slowing down of the TA branch in $\Gamma$-M direction.
The TO mode is the driver of the transition while the TA mode at finite $q$ in $\Gamma$-M direction largely corresponds to the motion of the primary order parameter - anti-phase Pb shifts. The generalized susceptibility corresponding to this order parameter in high-symmetry phase does not show any maximum in 3D reciprocal space on approaching the transition temperature, but the virtually whole $\Gamma$-M direction demonstrates a growth of the generalized susceptibility. In this case at the transition temperature the AFE order parameter is selected on the basis of the factors other than the harmonic susceptibility, namely on the basis of the anharmonic Umklapp interactions. By respectively repulsive and attractive biquadratic mode coupling the established anti-phase lead shifts suppress the ferroelectric instability and trigger the $R$-point oxygen octahedra rotations that correspond to the different order parameter.
A remarkable feature of the system addressed is that the ferroelectric instability, being the driving force for the condensation  of  the $\Sigma$-point order parameter, is suppressed by it, finally avoiding the formation of the ferroelectric state.

\section{Acknowledgments}
The authors would like to thank the Swiss National Science Foundation for funding this project.
The research leading to these results was also supported by the European Research Council under the EU 7th Framework Program (FP7/2007-2013) / ERC grant agreement [268058].
This research was supported by the Basic Science Research Program through the National
Research Foundation of Korea (NRF) funded by the Ministry of Education, Science and
Technology (2013R1A1A2006582) and by the National Centre for Science in Poland, grant 1955-B-H03-2011-40.
The Research was also supported  by Russian Academy of Sciences, Russian Foundation for Basic Research (Grants 14-02-01208 and 13-02-12429) and by grant of the government of the Russian Federation 2012-220-03-434, contract 14.B25.31.0025.
The IXS experiment was performed at the BL35XU of SPring-8 with the approval of the Japan Synchrotron Radiation Research Institute (JASRI) (Proposal No. 2011A1117, 2010B1497 and 2008B1240). The authors thank the the Swiss-Norwegian beam lines at ESRF for hospitality during experiment on diffuse scattering.
We would like to thank A. Bosak for sample preparation and for useful discussions on the interpretation of the results.
Prof. Sergei V. Kalinin is acknowledged for a useful discussion.

\bibliography{bibliography_PZOFlexoelectric}

\end{document}